\documentclass[preprint,12pt]{elsarticle}
\journal{Computer Physics Communications}

\usepackage{amsmath}
\usepackage[T1]{fontenc}
\usepackage{framed}
\usepackage{listings}
\usepackage[utf8]{inputenc}
\usepackage{url}
\usepackage{graphicx}

\usepackage{makecell}

\usepackage[dvipsnames]{xcolor}

\begin{document}

\begin{frontmatter}

\title{FIRE 7: Automatic Reduction with Modular Approach}

\author[SRCC,MCM]{Alexander V.~Smirnov\corref{cor1}}
\ead{asmirnov@srcc.msu.ru}

\author[HI]{Mao Zeng}
\ead{mao.zeng@ed.ac.uk}

\cortext[cor1]{Corresponding author}

\address[SRCC]{Research Computing Center, Moscow State University, \\ 119992 Moscow, Russia}
\address[MCM]{Moscow Center for Fundamental and Applied Mathematics, 119992, Moscow, Russia}
\address[HI]{Higgs Centre for Theoretical Physics, University of Edinburgh,\\
James Clerk Maxwell Building, Peter Guthrie Tait Road, Edinburgh, EH9 3FD,\\
United Kingdom}

\begin{abstract}
  FIRE7 is a major update to the FIRE program for integration-by-parts
  (IBP) reduction of Feynman integrals. A large part of improvements
  is related to the automatic reduction and reconstruction with the
  modular arithmetic approach, while the performance of the classical
  rational polynomial approach is also significantly increased. An
  improved presolve algorithm performs Gaussian elimination to
  simplify IBP identities before substituting numerical indices as in
  the Laporta algorithm. Various new command line tools are included
  to facilitate tasks such as applying an IBP reduction table to
  reduce a loop integrand as a linear combination of individual
  integrals.
\end{abstract}
\begin{keyword}
Feynman integrals \sep Integration by parts \sep Computer algebra \sep
Scattering Amplitudes \sep Perturbation theory \sep Finite-field arithmetic
\end{keyword}
\end{frontmatter}
\newpage

{\bf PROGRAM SUMMARY}

\vspace{1cm}

\begin{small}
\noindent
    {\em Program title:} FIRE, version 7 (FIRE 7)\\
    {\em CPC Library link to program files:} (to be added by Technical Editor) \\
    {\em Developer's repository link:} \url{https://gitlab.srcc.msu.ru/feynmanintegrals/fire.git} \\
    {\em Code Ocean capsule:} (to be added by Technical Editor)\\
    {\em Licensing provisions:} GPLv2\\
    {\em Programming language:} {\tt Wolfram Mathematica} 8.0 or higher, {\tt C++17}\\
    {\em Supplementary material:} see linked repository for installation instructions and examples\\
    {\em Journal reference of previous version:} https://doi.org/10.1016/j.cpc.2024.109261\\
    {\em Does the new version supersede the previous version?} Yes.\\
    {\em Reasons for the new version:} The new version brings major improvements to the performance of IBP reduction with modular arithmetic, enhanced parallelization, and a variety of convenient user tools.\\
    {\em Summary of revisions:} The new version enables MPI parallelization of finite-field IBP reduction runs and reconstruction of analytic results. Many performance improvements, especially those relevant for the modular arithmetic approach, are delivered. Linear combinations of Feynman integrals, e.g.\ as found in loop integrands, can be directly reduced. Various new utilities are provided for manipulating reduction tables.\\
    {\em Nature of problem:}
    The use of modular arithmetic is a major advance in efficient and parallelized reduction of Feynman integrals, and therefore a high-quality implementation as part of the FIRE program is desired. Many ``peripheral'' operations outside of the IBP reduction computation itself, such as the application of IBP tables to reduce a loop integrand, need efficient and convenient implementations.\\
    {\em Solution method:}
    Major new parallelized algorithms for finite-field IBP reduction
    are implemented. Gaussian elimination prior to seeding, i.e.\ a
    ``presolve'' step, enhances the efficiency of IBP
    reduction. Multiple numerical parameter points can be run
    simultaneously to significantly reduce resource overhead.
{\em References:} 
{\\}
[1] \url{https://home.bway.net/lewis/}, free--ware with some restrictions;
{\\}
[2] \url{https://doi.org/10.26089/NumMet.v24r425}, open source;
{\\}
[3] \url{https://flintlib.org/}, open source;
{\\}
[4] \url{https://symbolica.io/}, commercial software with free licenses for students and hobbyists

\end{small}

\section{Introduction}

The evaluation of Feynman integrals is a fundamental component of perturbative quantum field theory computations. A key technique in this process is integration-by-parts (IBP) reduction \cite{Chetyrkin:1981qh, Tkachov:1981wb}, which allows all integrals within a given family (characterized by a set of propagators) to be expressed in terms of a finite set of simpler, so-called master integrals. This method leverages the fact that total derivatives vanish under dimensional regularization, giving rise to linear relations between integrals, known as IBP identities, which are often supplemented by Lorentz invariance relations \cite{Gehrmann:1999as}. Typically, the master integrals are chosen for their relative simplicity, involving low-degree numerators and propagators raised to low powers.

In all but the most basic cases, performing reduction requires combining IBP identities in nontrivial ways. A systematic procedure for this is the Laporta algorithm \cite{Laporta:2000dsw}. Several public implementations of this algorithm exist, including {\tt AIR} \cite{AIR}, {\tt Reduze} \cite{Studerus:2009ye, Reduze}, {\tt LiteRed} \cite{Lee:2012cn, Lee:2013mka, LiteRed2}, {\tt FIRE} \cite{Smirnov:2008iw, Smirnov:2013dia, Smirnov:2014hma, FIRE6}, and {\tt Kira} \cite{Maierhofer:2017gsa, Kira, Klappert:2020nbg}. The authors of this paper have been developing {\tt FIRE} for years, a program that has been widely used in state-of-the-art computations in high-energy physics. The Laporta algorithm works by generating all IBP identities up to a specified complexity, followed by solving the resulting system of linear equations via Gaussian elimination. These equations have polynomial coefficients in the spacetime dimension $d$ and in kinematic parameters such as Mandelstam variables and masses, which turn into rational functions during Gaussian elimination. To avoid a blow-up in expression size, it is critical to simplify these rational functions~---~typically by reducing them to a single fraction with a co-prime numerator and denominator throughout all the reduction steps.

Alternative strategies to the Laporta algorithm include deriving symbolic reduction rules that depend on propagator powers and irreducible scalar products. Such rules can be obtained using heuristic methods (as in {\tt LiteRed} \cite{Lee:2012cn, Lee:2013mka}), techniques involving Gröbner bases \cite{Tarasov:2004ks, Gerdt:2005qf, Smirnov:2005ky, Smirnov:2006tz, Smirnov:2006wh, Lee:2008tj, Barakat:2022qlc}, generating functions ~\cite{Guan:2023avw, Hu:2024kch, Feng:2025leo}, and variants \cite{Kosower:2018obg, Smith:2025xes} of the syzygy equation method also mentioned later. These rules reduce more complicated integrals to simpler ones according to a prescribed ordering, and repeated application is usually required to fully reduce an integral to master integrals. Again, rational function simplification plays a vital role in managing intermediate expression complexity. Methods based on syzygy equations \cite{Gluza:2010ws, Schabinger:2011dz, Ita:2015tya, Larsen:2015ped, Abreu:2017xsl, Abreu:2017hqn, Bohm:2018bdy, Bendle:2019csk, Wu:2023upw, Wu:2025aeg, Page:2025gso} aim to shrink the size of the linear system, while techniques employing block-triangular forms \cite{Liu:2018dmc, Guan:2019bcx, Guan:2024byi} help organize the system more efficiently. There is also work on bypassing the need to solve linear systems of IBP equations \cite{Mastrolia:2018uzb, Frellesvig:2020qot}. Nevertheless, the simplification of rational functions remains a central computational challenge in all these approaches.

An alternative approach to reduction involving rational functions of many variables is performing numerical evaluations followed by analytic reconstruction. However, floating-point arithmetic is unsuitable due to the need for exact coefficients, and rational arithmetic—though exact—requires costly arbitrary-precision computations. To overcome this, Refs.\ \cite{Kant:2013vta, vonManteuffel:2014ixa, Peraro:2016wsq} proposed using finite field arithmetic for IBP reduction. Finite fields offer significant efficiency advantages, as elements are represented by machine-size integers with exact arithmetic. Although finite fields entail information loss, this can be resolved by reconstructing rational numbers from evaluations in several fields using the Extended Euclidean Algorithm \cite{Wang1981,Monagan2004}, based on the Chinese Remainder Theorem \cite{Gathen2013}.

Once numerical data are obtained, the main task is reconstructing the target function (the "black box") from its sampled values. Univariate interpolation techniques such as Newton and Thiele methods \cite{AbrSte64} are classical, but multivariate reconstruction is more recent. Sparse multivariate interpolation—--assuming a bounded number of terms—--has been developed in Refs.\ \cite{Zippel79,BenOrTiw88,Grigoriev1994,Kleine2005,HomoMulti2011a,HuaGao17}. Dense reconstructions, without such bounds, are less common. A Thiele-based generalization was implemented in FIRE6 \cite{Smirnov:2019qkx}, but it struggles beyond two variables. The most effective multivariate approach is based on homogeneous scaling \cite{HomoMulti2011a}, adopted in {\tt FiniteFlow} \cite{Peraro:2019svx} and {\tt FireFly} \cite{Klappert:2019emp}, the latter integrated with Kira 2.0 \cite{Klappert:2020nbg} and Kira 3.0 \cite{lange2025kira3integralreduction}. In~\cite{Belitsky:2023qho} we suggested a balanced reconstruction approach as an alternative to homogeneous scaling, later combined with the Zippel algorithm to take advantage of sparsity \cite{Smirnov:2024onl}, which is integrated into this new version of {\tt FIRE}.

In this paper, we present a new version of {\tt FIRE} with automates the modular arithmetic approach, including reduction and subsequent reconstruction, ``with one command'' with MPI parallelization. In section~\ref{sec:modular} we describe our approach. In section~\ref{sec:nonmodular} we describe other new improvements in {\tt FIRE7}. In section~\ref{sec:usage} the installation and usage of {\tt FIRE} is presented. In section~\ref{sec:exampleAndBenchmark} we provide examples and benchmarks.

\textbf{Note:} for the rest of the paper, unless otherwise specified, file locations are relative to the folder {\tt fire/FIRE7}.

\section{Modular approach}

\label{sec:modular}

\subsection{Overview of the approach}

One of the most important features presented in the new version of FIRE is the ability to automatically perform reduction with the modular approach\footnote{To disambiguate, ``modular'' always refers to modular arithmetic in this paper.} with the subsequent reconstruction of coefficients, i.e.\ inferring analytic expressions from a large number of exact numerical evaluations in finite fields. With this feature, we aim to assist the user as much as possible with reconstruction, especially in the case with multiple variables, where a manual reconstruction process would become very cumbersome requiring multiple reconstruction sub-steps (Thiele reconstruction, Newton reconstruction etc.) proceeding with variables one by one.

This approach relies on the following parts of the FIRE package:
\begin{itemize}
 \item The FIRE reduction program itself, specifically the {\tt bin/FIRE7p} binary (where {\tt p} refers to the prime variant) which performs reduction in a single probe point, i.e. a set of values given to space-dimension and kinematic invariants and a large prime number chosen for reduction, or alternatively the {\tt FIRE7mp} binary (where {\tt mp} refers to the multiprime variant) which simultaneously performs reductions in multiple probe points;
 \item The {\tt bin/reconstruction} tool which takes some tables generated by FIRE and creates a reconstructed table (each call performs either reconstruction for the analytic dependence on one more variable or rational reconstruction, i.e.\ recovering rational numbers from finite-field images.);
 \item An MPI program {\tt bin/FIRE7$\_$MPI} that calls both the FIRE reduction program and the reconstruction tool in order to obtain the reconstructed result automatically.
\end{itemize}

The reason for using such an approach was discussed many times in papers on Feynman integrals. The modular approach is an alternative to the classical (rational polynomial) approach where the reduction is performed with coefficients being rational functions of many variables. Within the classical approach, almost all the time consumed by the program is spent on simplifying coefficients with an external program. (In the past {\tt fermat} was the dominant choice, but now the {\tt FLINT} library has proven to be more effective and is adopted as the default in FIRE.) The time needed for reduction as well as RAM required can be too large to fit into a reasonable session, so there is always a risk of running out of RAM or encountering a hardware failure, making one lose months of calculation if it is interrupted in an unexpected state.

The modular approach is much more resistant to possible interruptions since one might need millions of intermediate reduction runs, each of which results in a separate table file stored on disk. Moreover, the modular approach can efficiently make use of multiple machines working on the same task, especially in the case of supercomputers. However, the MPI-parallelized program is not restricted to clusters and can be launched on a single machine as well.

\subsection{Modular reduction in FIRE}

\label{sec:mod_rec}

To manually call modular reduction with {\tt FIRE7p} one should create a {\tt config} file as usual (see the full {\tt config} syntax in Section \ref{sec:config}), and then {\tt FIRE7p} should be called with the {\tt -{}-variables} option. The variable values should be underscore-separated, and the last position is the index of the large prime number used as the finite-field modulus. The primes used by {\tt FIRE} are stored in the {\tt sources/tools/primes.cpp} file, where at position $0$ we use the number $2017$ (purely for testing purposes), followed by large primes, $127$ of which are close to $2^{64}$ at positions $1$ to $127$ and are recommended for regular use as the modulus. The first six primes in this list are:
18446744073709551557, 18446744073709551533, 18446744073709551521,\\
18446744073709551437, 18446744073709551427, 18446744073709551359.\\
There are also $128$ primes close to $2^{63.5}$ at positions $128$ to $255$ and $128$ primes close to $2^{31}$ at positions $256$ to $383$. Variable values can be numbers or the $b^e$ syntax for a given base $b$ and exponent $e$. It is also possible to automatically use the prime numbers at positions $128$ onward as base values for variables, using the notation $p0$, $p1$, \dots, denoting prime numbers from the {\tt primes.cpp} file at positions $128$, $129$, \dots (note the offset). Here are some valid examples:

\begin{itemize}
\item \verb|bin/FIRE7p --variables 80_90_100_1 --config |\\\verb| examples/doublebox3N --parallel|
  
  This uses $t=80, s=90, d=100$, with the modulus as the prime at position 1, 18446744073709551557.
 \item \verb|bin/FIRE7p --large_variables --variables 80^1_90^1_100^1_1 |\\\verb| --config examples/doublebox3N --parallel|
  
  This is the same as the previous one, as the exponents for the variables are 1. The {\tt -{}-large-variables} option is required for using the exponent notation, as large intermediate numbers can be produced by exponentiation before modding by the prime modulus.
 \item \verb|bin/FIRE7p --variables p1^5_p2^6_p3^7_2 --config |\\\verb| examples/doublebox3N --parallel|
  
  This uses $t=13043817825332782193^5, s=13043817825332782171^6, d=13043817825332782093^7$, where the base values are primes at positions 128, 129, and 130, with the modulus as the prime at position 2, 18446744073709551533.
\end{itemize}

{\tt FIRE7p} saves reduction results in {\tt tables} files. Since one is going to call {\tt FIRE7p} multiple times, the output table filenames are different~---~before the final {\tt.tables} extension, {\tt FIRE7p} appends the variable values as supplied by the {\tt -{}-variables} option. For example, with the second variant above, the table filename will be \verb|doublebox_80^1_90^1_100^1_1.tables| assuming that the {\tt config} file has specified the output filename \verb|doublebox.tables|.

\subsection{Reconstruction tool}

\label{sec:reconstruction}

The reconstruction tool {\tt bin/reconstruction} takes a number of tables, performs reconstruction of all coefficients, and saves the results as a new table file. For most users, the MPI wrapper covered in Section \ref{sec:mpiWrapper} automates the reconstruction steps in a more convenient manner, while internally calling {\tt bin/reconstruction}, so this subsection can be skipped for a first reading. For users who want more manual control, e.g.\ generating tables and performing reconstructions step-by-step, {\tt bin/reconstruction} can be used directly, assuming familiarity with the balanced Zippel reconstruction algorithm \cite{Belitsky:2023qho, Smirnov:2024onl}.  There is a number of reconstruction methods this tool supports, chosen by the {\tt -{}-method} command line option:

\begin{itemize}
 \item {\tt rational}: rational reconstruction from finite-field images to rational numbers. It is advisable to have rational reconstruction as the final reconstruction step after all other reconstructions, as explained in e.g.~\cite{Belitsky:2023qho};
 \item {\tt thiele}: Thiele reconstruction by one variable; in practice should be used only for coefficients not depending on other variables, i.e.\ for the first variable and to create balancing tables \cite{Belitsky:2023qho, Smirnov:2024onl};
 \item {\tt newton}: newton reconstruction by one variable; implemented but not used in pure form by {\tt FIRE7\_MPI};
 \item {\tt numeratorNewton}: assuming the last reconstruction variable is the space-time dimension $d$ and masters are chosen in a proper way \cite{Smirnov:2020quc, Usovitsch:2020jrk}, we know that the denominators of all master coefficients can be factorized into a function of $d$ and a function of other (kinematic) variables. Thus, after a Thiele reconstruction by $d$ for one set of values of other variables, the $d$-dependent part of the denominator is known, and further reconstructions can be performed with this knowledge. This is because knowing the denominator in $d$, one can multiply by it and use Newton reconstruction for the numerator when knowing the reconstruction value in the kinematic variables;
 \item {\tt balancedNewton}: the method used to proceed from $k$ reconstructed variables to $k + 1$ with the use of balancing tables so that the numerators and denominators are reconstructed independently~\cite{Belitsky:2023qho};
 \item {\tt balancedZippel}: having a reconstructed function of $k$ variables for one value of the subsequent variable, this method reconstructs the function for $k$ variables and a different value of the subsequent variable; not used directly by {\tt FIRE7\_MPI};
 \item {\tt balancedZippelNewton}: the most efficient method to go from $k$ reconstructed variables to $k + 1$ reconstructed variables (especially for multiple variables and thus sparse coefficients), equivalent to reconstruction with {\tt balancedZippel} followed by {\tt balancedNewton} without storing intermediate results.
 \end{itemize}

 An experimental Mathematica interface to the reconstruction tool is\\
 {\tt mm/FireReconstruct.wl}. The Mathematica notebook\\
 {\tt examples/manualReconstructionExample.nb}\\
 explains the usage and
 demonstrates how one could use the reconstruction tool for general
 purposes (not necessary for IBP).  The notebook transparently shows
 the shell commands called under the hood, which can also help with
 learning the manual usage of the reconstruction tool.

\subsection{MPI wrapper}
\label{sec:mpiWrapper}
The so-called MPI wrapper, the binary {\tt bin/FIRE7\_MPI} is a program that is designed to automatically run modular reduction and subsequent reconstructions. The full launch syntax will be described later, and here we give a general overview. First of all, this program uses MPI, i.e.\ Message Passing Interface, a protocol that allows communication between multiple processes that can be launched on different computers at the same time. A native way to launch the program locally is \\
\verb|mpirun -np <n_processes> bin/FIRE7_MPI <arguments>|\\
where {\tt <n\_processes>} should be replaced with the number of processes to run concurrently. (When running on one local computer, typically one can set it equal to the number of cores or exceed it by one since the master process does not utilize CPU much). When running on a cluster with multiple machines for one reduction, one should refer to the local cluster documentation on how MPI programs are launched. The basic syntax to call the wrapper is, e.g.\\
\verb|mpirun -np 5 bin/FIRE7_MPI -Z --reconstruct -P 3 -I \|\\\verb| 80_90_100 -S -c examples/doublebox3N|\\
Here we provide starting variable values $t=80, s=90, d=100$, use the Zippel approach with the {\tt -Z} flag, ask the code to do the reconstruction and set a limit of a maximum of $3$ prime values. Also we pass the important information that the last variable can be split (factorized) in denominators \cite{Smirnov:2020quc, Usovitsch:2020jrk} with the {\tt -S} flag. This call is sufficient to initiate IBP reduction computations at numerical parameter values, and the run completes when the desired reduction table is reconstructed.

The order in which the variables are placed is very important. As a starting point, we suggest running {\tt FIRE} with any order and to detect Thiele ``limits''~---~the number of probe points to perform a one-dimensional Thiele reconstruction by the corresponding variable. We recommend reordering the variables in the decreasing order of Thiele limits, i.e.\ staring from variables that need a large number of probe points in Thiele reconstruction. However, if one can chooses a basis of master integrals such that the denominators by $d$ are split, it is recommended to put $d$ at the last place, as the smaller Newton limit will be used for it instead.

\subsection{Reducing for multiple numerical probes simultaneously}
\label{sec:multiprime}
Working within the modular approach has its disadvantages. One of those is that the overhead for solving the linear system of equations becomes too large compared with the actual solving, i.e.\ performing modular arithmetic operations during Gaussian elimination. To decrease the overhead and also to decrease the number of table files stored on disk, we created a FIRE variant that performs reduction at multiple probe points at the same time. By default {\tt FIRE7} uses $16$ points in this variant, and this can be changed by the {\tt configure} script which requires a full rebuild of {\tt FIRE}.

To use such a reduction program one should call the {\tt bin/FIRE7mp} binary (where {\tt m} stands for multiple) with almost the same syntax, but adding a plus symbol at the places where the variables differ. The following examples are valid:

\begin{itemize}
\item \verb|bin/FIRE7mp --variables 80_90+_100+_1 --config |\\\verb| examples/doublebox3N --parallel|

  Here, $80$ stays as the constant value for the first variable, and the second and third variables get shifted in pairs $(90, 100), (91, 101), \ldots, (105, 106)$.
  
\item \verb|bin/FIRE7mp --variables 80^1_90^1+_100^1+_1 --config |\\\verb| examples/doublebox3N --parallel|

  Here, $80$ stays constant for the first variable, and the second and third variable exponents get shifted in pairs $(90^1, 100^1)$, $(90^2, 100^2)$, \dots, $(90^{16}, 100^{16})$.
\end{itemize}

The resulting table files also have the plus symbol at the places specified in the variables setting, to indicate that IBP reduction results for multiple parameter values are stored in the same file. {\tt FIRE7\_MPI}, when using {\tt FIRE7mp} with the {\tt -M/--multitables} option, understands the above table file format during automated reconstruction. Please note that the number of values is not saved in the table filename, so in case {\tt FIRE} is reconfigured for another value for the number of probe points, the old tables cannot be used.

\subsection{Reduction and reconstruction of combinations}

In some situations, e.g.\ when working with differential equations, one does not need reduction results for particular Feynman integrals but their linear combinations.
It is also known that the master coefficients in the reduction results of these combinations can be much simpler than the ones for individual integrals involved.
To reduce a combination, one basically needs to reduce all integrals in it, so with the classical analytic IBP approach, working with combinations does not help much.
However, with the modular approach, the use of combinations becomes efficient. The reason is that we can run the reduction for individual integrals but save only tables for combinations.
Thus, the coefficients to be reconstructed become much simpler, resulting in fewer required probe points for the reconstruction.

The syntax for specifying combinations can be found in {\tt examples/boxc.m}.

\section{Improvements not related to the modular approach}

\label{sec:nonmodular}

There is a number of improvements for the operation of FIRE in general, not exclusively for the modular approach. This section summarizes the most important features. Some of them already appeared in the public release of version {\tt FIRE6.5}, but the features themselves were not documented at that point, so we present them here. Moreover, some of them were updated and improved after the initial inclusion in the {\tt FIRE6.5} code.

\subsection{Split masters mode}
\label{sec:split}
An alternative to the modular approach for the parallelization of IBP reduction is performing separate computations of the coefficients of different master integrals within the classical (rational polynomial) approach. To proceed this way, one needs to determine first the list of master integrals, for example, by running a sample modular job for a single probe point. Afterwards, there is a syntax to use {\tt FIRE7} so that only a subset of master integrals is considered to be non-zero, and other ones are directly set to zero in the system of IBP linear equations. This simplifies the reduction and reduces the consumption of RAM and CPU time.

The {\tt \#masters |first-last|filename} option in the configuration file enables this mode. Here, {\tt first} and {\tt last} are numerical positions of master integrals listed in the following file (the indexing starts from 1 not 0). For example, {\tt \#masters |1-2|doublebox.masters} sets all master integrals, other than the 1st and 2nd ones listed in {\tt doublebox.masters}, to zero. Similarly, {\tt \#masters |4-4|doublebox.masters} retains only the 4th master integral. This allows parallel runs that each targets a subset of master integrals. One should be sure that no more master integrals are going to appear, otherwise {\tt FIRE} will crash immediately upon encountering one. The output table files will have a suffix with the master range {\tt first-last} before the {\tt .tables} extension. After all reductions (targeting different subsets of master integrals) are ready, one can combine them in {\tt Mathematica} with the {\tt CombineTables} function. The syntax is {\tt CombineTables[filename, list]}, for example, {\tt CombineTables[filename, \{\{1, 2\}, 3, 4\}]}. The list can contain either numbers or pairs of numbers indicating the start and end of a range.

It might also be useful to provide the list of masters, with the syntax {\tt \#masters filename}, even if the split mode with the extra {\tt |first-last|} syntax is not used, as this knowledge can improve reduction efficiency.

For historical reasons, the {\tt \#masters} option also has another effect when used without any additional argument such as the filename. In this case, FIRE will run in a different mode which only finds the list of master integrals but not does not perform the backward step of the IBP reduction.
\subsection{Presolving IBP relations}

\label{sec:presolve}

There are two opposite approaches to working with IBP equations. The initial one, originating from Laporta, is to generate all needed relations with symbolic indices (propagator/ISP power) and to solve the system after substituting specific numerical values of the indices. The other one aims to solve relations before substituting indices, i.e.\ keeping symbolic dependence on the indices, which was attempted with the use of Gröbner bases \cite{Gerdt:2004kt, Smirnov:2005ky, Smirnov:2006wh, Smirnov:2006tz}, {\tt LiteRed} \cite{Lee:2012cn, Lee:2013mka, LiteRed2} by Roman Lee, generating functions \cite{Guan:2023avw, Hu:2024kch}, and variants of the syzygy equation method \cite{Kosower:2018obg, Smith:2025xes}. However, in many cases, trying to solve all relations without substituting indices fails or makes the relations too complex.

The pre-solving of IBP relations is an attempt to somewhat combine these approaches and to allow the reduction program to combine IBP identities before substituting indices, but only linearly, that is, not applying shift operators to them. This reduces the complexity of the system without making the equations have too many terms.

The option is on by default and can be switched off with the option {\tt \#no\_presolve} in the {\tt config} file or reverted to the variant implemented in {\tt FIRE6.5} by {\tt \#old\_presolve}.
The old presolve variant takes IBPs relations and performs forward Gaussian elimination on them, aiming at positive shifts in the chosen sector. The new full presolving also applies backward Gaussian elimination.
We expect the full presolve to be efficient in most cases, but are aware of situations where one might wish to switch it off or use the old variant.

\subsection{Orderings}

\label{sec:orderings}

As suggested in~\cite{Klappert:2020nbg}, ordering choices in solving the IBP linear system can have a great impact on performance. {\tt FIRE} now allows users to set the ordering choices as a sequence of letters in the {\tt config} file. The ordering is generated in each sector, taking absolute differences of integral indices and the sector corner (i.e. the index vector where positive powers are replaced by 1 and non-positive powers are replaced by zero). The ordering letters are applied in the order they appear.

The following letters are valid:

\begin{itemize}
 \item {\tt A}: $1$ at each position, i.e. sum of all index shifts;
 \item {\tt P}: sum of all positive index shifts;
 \item {\tt N}: sum of all negative index shifts;
 \item {\tt r}: modifier for the next letter such as {\tt a, p, n} indicating that the ordering part will be reversed (from last to first);
 \item {\tt i}: modifier for the next letter such as {\tt a, p, n} indicating that the ordering part will be inverse lexicographic, i.e.\ swapping ``less than'' and ``greater than'' of lexicographic ordering;
 \item {\tt a}: lexicographic ordering over all indices;
 \item {\tt p}: lexicographic ordering over positive indices;
 \item {\tt n}: lexicographic ordering over negative indices;
 \item {\tt l}: special ordering originating from papers of Lee~\cite{Lee:2008tj}.
\end{itemize}

The use of a proper ordering setting can have a significant impact on performance.

\subsection{Preferred integrals}

\label{sec:pos_pref}
Although this is not exactly a new feature, it requires an explanation since it might often confuse users. {\tt FIRE} automatically detects master integrals as those integrals in sectors that cannot be reduced to lower sectors or lower integrals in the same sector under the chosen ordering. Sometimes, one needs to influence the choice, and this can be done with the preferred integral list and the {\tt pos\_pref} setting. Note that in case a sector has integrals provided with {\tt \#preferred}, then the {\tt pos\_pref} has no impact on that sector.

The preferred integral list makes {\tt FIRE} mark some integrals in a sector with a preferred bit (in addition to data fields storing the indices in the C++ data structure representing the integrals), making them lower than all unmarked integrals in the same sector (under the comparison function used in the code). This however does not guarantee that that they are going to become master integrals, if only because one might specify more preferred integrals in the sector than there are masters in it. Unlike the {\tt master} setting, specifying extra preferred integrals does not cause the reduction to crash. All integrals listed as preferred in a sector are compared between each other under the chosen ordering, which determines which subset will be made master integrals. It is also important to mention that all integrals listed as masters and all integrals on the left-hand sides, and right-hand sides of provided rules using the {\tt rules} setting are also treated as preferred.

Another option related to preferred integrals is the {\tt pos\_pref} option of the configuration file. By default, it is equal to $1$. This option is applied to sectors that have no manually listed preferred integrals and makes all integrals with up to {\tt pos\_pref} dots preferred, meaning by default {\tt FIRE} prefers integrals with $1$ dot as masters to integrals with $1$ irreducible numerator no matter what ordering is chosen. This is the typical setting when the user tries to choose masters with dots. However, in some sectors there might be many masters, and there are not enough integrals with $1$ dot for this, so the {\tt pos\_pref} can be increased to ensure that integrals with dots, rather than numerators, are chosen as masters.

Also, {\tt pos\_pref} can be made negative to prefer masters with numerators up to a rank equal to the the absolute value of {\tt pos\_pref}.

Another use of {\tt pos\_pref} is together with {\tt LiteRed}~---~{\tt FIRE} can use internal sector symmetries of {\tt LiteRed} and apply them to preferred integrals. The reason to not apply them to all integrals is that this sometimes slows down the reduction.

\section{Installation and usage}

\label{sec:usage}

In this section, we aim to provide a guide on how to install and use {\tt FIRE} listing important options. The previous manual \cite{FIRE6} was published more than five years ago and definitely needs an update.

\subsection{Installation}

To install {\tt FIRE7} one should first clone it from the gitlab repository:

\noindent {\tt git clone https://gitlab.srcc.msu.ru/feynmanintegrals/fire.git}

One can also add one more argument to the clone command specifying the folder ({\tt fire} by default).

Note that there is a set of required common libraries and compiler tools not shipped with {\tt FIRE}. The most common set on a Debian-based distribution is\footnote{{\tt libcln-dev} can be omitted unless one wants to use the non-default {\tt ginac} backend for simplifying rational functions.} \\
{\tt apt-get install g++ make cmake automake \textbackslash \\autoconf zlib1g-dev libcln-dev}

Other commands should be performed in the {\tt fire/FIRE7} folder of the newly cloned repository. The first step is to launch the {\tt configure} script by {\tt ./configure} providing options if needed. There are the following options:

\begin{itemize}
 \item {\tt -{}-enable-debug} --- enables debugging symbols and stack trace printing in case of crashes. Has minimal impact on performance;
 \item {\tt -{}-mprime=value} --- sets a non-default number of probe points used in {\tt FIRE7mp}; please note that the tables generated with different values are incompatible;
 \item {\tt -{}-enable-zlib, -{}-enable-zstd, -{}-enable-snappy} --- enable compressors which can be set in {\tt config} files;
 \item {\tt -{}-cpp=value, -{}-cc=value} --- set {\tt c++} and {\tt c} compilers, defaults are {\tt g++} and {\tt gcc};
 \item {\tt -{}-enable-tcmalloc, -{}-enable-mimalloc} --- two options to choose another memory allocator (incompatible with each other);
 \item {\tt -{}-build-openmpi} --- to build the openmpi compiler in case the one on the cluster is too old;
 \item {\tt -{}-prebuilt-mpfr, -{}-prebuilt-gmp, -{}-prebuilt-flint,} and \\ {\tt -{}-prebuilt-zstd} --- use a prebuilt version of listed libraries; this shorten compile times but might be incompatible with the system in use;
 \item {\tt -{}-enable-ginac, -{}-enable-pari, -{}-enable-symbolica} --- used to enable alternative simplifying libraries (by default {\tt FIRE} comes with {\tt FLINT} and {\tt Fermat}); note that {\tt Symbolica} \cite{Symbolica} needs a valid license;
 \item {\tt -{}-fermat7} --- use version $7$ instead of $5$ of {\tt Fermat};
 \item {\tt -{}-enable-lto} ---  enable gcc link time optimization for the prime version. This can result in a 5 percent speedup, but spoils debugging and crash diagnostics;
 \item {\tt -{}-small-point, -{}-many-sectors[=value], -large-point} --- these options influence how large the point size (the amount of memory for storing the indices of an individual integral) is, and therefore the maximal number of indices and how many non-zero sectors one can have; the possible variants are shown in Table \ref{tab:pointsizes}.
\end{itemize}

\begin{table}[ht]
\centering
\begin{tabular}{lcccc}
\hline
\textbf{Options} & \textbf{Point size} & \textbf{Max. shift} & \textbf{Max. value} & \textbf{Sectors} \\
\hline
\texttt{-{}-small-point}       & 16 & 18 & 31 & $2^{15}$ \\
default                      & 24 & 22 & 127 & $2^{15}$ \\
\texttt{-{}-many-sectors}      & 24 & 20 & 127 & $2^{20}$ \\
\texttt{-{}-many-sectors=\textit{n}} & 24 & 20 & 127 & $2^{n}$ \\
\texttt{-{}-large\_point}      & 32 & 28 & 127 & $2^{20}$ \\
\texttt{-{}-large\_point} and     &  &  &  &  \\
\texttt{-{}-many-sectors=\textit{n}}      & 32 & 28 & 127 & $2^{n}$ \\
\hline
\end{tabular}
\label{tab:pointsizes}
\caption{Different point combinations}
\end{table}

The point size option might be important, especially in the modular case, since each stored coefficient uses only $8$ bytes, and the point size might greatly influence the amount of RAM needed. However, the significance is decreased in the multiprime mode. Max.\ shift stands for the maximal total of the absolute differences between indices of integrals involved and the corresponding sector corners. For example, the shift of $(1, 3, 5, -2)$ is $(1-1) + (3 - 1) + (5 - 1) + (-(-2)) = 8$. For diagrams up to $5$ loops, $22$ indices should normally be sufficient, while for $6$ loops, one should switch to a larger point size. Be careful with the small point setting to make sure that the total shift of indices does not exceed a relatively small number (max.\ shift). The ``Sectors'' column of the table stands for the number of non-zero sectors; with the \texttt{-{}-many-sectors} option, $4$ bytes are used to store a sector number---but do not blindly increase the setting, as there is a proportional amount of RAM needed to store sector numbers and some pointers---e.g.\ with \texttt{-{}-many-sectors=31}, around $12$ GB will be used soley for this.

An attentive user might also notice the {\tt -{}-enable-fire7np} which enables private modular reduction improvements as a git submodule that is not yet made public.

After running the {\tt configure} script, the options it received are saved in the {\tt previous$\_$options file}, and a {\tt Makefile} is created. Now one can build the libraries that {\tt FIRE} depends on with the {\tt make dep} command. The more options enabling libraries above were set, the more libraries are going to be built at this point. As usual, {\tt make dep -j} or {\tt make dep -j \textit{n}} \textit{n} will enable building libraries in parallel. However, there is a known issue that some of the libraries on some filesystems refuse to build in parallel if they cannot lock some file, so a non-parallel build has to be used.

After the libraries are built properly, one can build {\tt FIRE} with {\tt make}. If the MPI part is needed, also {\tt make mpi} should be called. The build can be verified with {\tt make test}.

An alternative way to build {\tt FIRE} is to use Docker, which simplifies the installation process and runs FIRE in a self-contained sandbox. The FIRE image can be obtained from Docker Hub with the command\\
{\tt docker pull asmirnov80/fire}\\
Now one can launch {\tt FIRE} from the container like\\
{\tt docker run -{}-rm fire /app/FIRE7/bin/FIRE7 -{}-help}\\
As Docker containers are sandboxed, any host folder needs to be explicitly mounted as a volume to be accessed by FIRE inside the container. For example, the following command mounts the current directory as the {\tt /host} folder in the container and makes it the working directory when launching FIRE:\\
{\tt docker run -{}-rm -v \$\{PWD\}:/host -w /host fire \textbackslash \\
  /app/FIRE7/bin/FIRE7 -{}-help}\\

For advanced users who want to re-build the Docker image after local modification to the FIRE code, here are the instructions. First, make a new clean clone of {\tt FIRE}, change the directory to the top-level {\tt fire} folder and run\\
{\tt docker build -t fire .}\\
to build the image.\footnote{\textit {Warning}: trying to build a docker image within a folder where {\tt FIRE} has been already built might lead to build errors or unexpected behavior, since each time this command is issued, the content of the {\tt FIRE7} folder is copied into the container and rebuilt.}
Currently the build uses the \\
{\tt https://hub.docker.com/repository/docker/asmirnov80/for$\_$fire/} \\image with preinstalled external libraries as a builder image to bootstrap the final image. The builder image in turn uses the {\tt debian:trixie-slim} base image.

\subsection{Configuration file settings}
\label{sec:config}
Since the {\tt Mathematica} part of {\tt FIRE} remains almost unchanged since previous versions, we do not repeat those instructions and instead refer to the previous manual. On the other hand, the {\tt c++} part has changed a lot, so here we will list the possible commands in the {\tt config} file, the main file for launching the {\tt c++} reduction. As a reminder, having a {\tt config} file with the name {\tt example.config}, one launches {\tt FIRE} with {\tt bin/FIRE7 -c example}, and the {\tt .config} extension is added automatically when FIRE reads the file. The configuration file supports a number of instructions, each of which begins with the symbol {\tt \#}. Double {\tt \#\#} is a commented line and has no effect. The order of some commands is important, but the current version is supposed to stop immediately in case of an incorrect order.

The first group of commands should be placed before the {\tt \#start} directive:

\begin{itemize}
 \item {\tt \#calc}: sets the simplifier library, default is {\tt flint} (in previous versions, the default was {\tt fermat});
 \item {\tt \#ordering}: an important setting specifying an ordering type, detaild in section~\ref{sec:orderings}; the default value according to the notation in that section is {\tt ANl};
 \item {\tt \#index\_ordering}: specifies order of indices in the ordering, by default from first to last;
 \item {\tt \#positive}: a way to provide comma-separated indices that
   have to be positive, i.e.\ IBP on a cut (see e.g.\ Ref.\ \cite{Larsen:2015ped}); a fast way to perform partial reduction;
 \item {\tt \#compressor}: sets compressor for coefficients, default is {\tt lz4}, can be switched off with {\tt none};
 \item {\tt \#threads}: sets the number of threads working in parallel, each thread in its own sector (of the same level, i.e.\ the number of propagators);
 \item {\tt \#sthreads}: overrides the threads setting at the substitution stage;
 \item {\tt \#fthreads}: sets the number of simplifier library threads shared by evaluation threads. If the number is preceded by the $\tt s$ letter, then each thread has the specified number of its own simplifier threads;
 \item {\tt \#variables}: lists comma-separated variables that exist in expressions; variables can be followed with {\tt ->} and a value;
 \item {\tt \#database}: provides a path to temporary database files;
 \item {\tt \#storage}: provide a path to another folder where database files are copied so they are not broken in case of job interrupts; useful to be able to continue a job from an intermediate point;
 \item {\tt \#forward} and {\tt \#backward}: run only forward (first stage) or backward (second stage) run of {\tt FIRE}; needs the {\tt \#storage} option to be used;
 \item {\tt \#bucket}: fine-tuning of kyotocabinet database;
 \item {\tt \#wrap}: makes databases wrapped into a single file; might be useful in order to decrease the number of files on disk;
 \item {\tt \#clean\_databases}: cleans up temporary databases after completing the reduction;
 \item {\tt \#prime}: for the modular version, specifies the prime number index (but it is more advisable to provide it from the command line);
 \item {\tt \#allIBP}: makes all IBPs used, switches off the Lee ideas \cite{Lee:2008tj} to remove some of those;
 \item {\tt \#no\_presolve} and {\tt \#old\_presolve}: see section \ref{sec:presolve};
 \item {\tt \#even} and {\tt \#odd}: make {\tt FIRE} consider all integrals of other parity equal to zero;
 \item {\tt \#pos\_pref}: see the \ref{sec:pos_pref} section.
\end{itemize}

After the options listed above, the {\tt \#start} directive should appear, followed by instructions related to file paths of the diagram in use:

\begin{itemize}
 \item {\tt \#folder}: provides a path to other files so that full path can be written in one place;
 \item {\tt \#hint}: an experimental setting providing a path to a folder where ``hint'' files are saved. A first run saves the hint files, and all subsequent runs with access to the hint files might use less time knowing which IBP relations are to be used. It might make sense to create hint files with a modular run before a polynomial run. However, this option has not been tested well, and we are currently unable to provide proper benchmarks to confirm its usefulness. Please use it at your own risk;
 \item {\tt \#rules}: path to a file with rules --- mappings of individual integrals to linear combinations of other integrals; rules cannot be recursive;
 \item {\tt \#lbases}: path to a file with rules originating from {\tt LiteRed};
 \item {\tt \#output}: path to output file with tables;
 \item {\tt \#preferred}: path to a file specifying preferred master integrals;
 \item {\tt \#integrals}: path to input file with integrals.
\end{itemize}

\subsection{Command-line options of FIRE}

The main {\tt FIRE} binaries ({\tt bin/FIRE7}, {\tt bin/FIRE7p}, {\tt bin/FIRE7mp}) also accept some of the options from the command line. The explanations are available if calling {\tt bin/FIRE7 -{}-help}, most of the options correspond to the configuration file directives. Let us outline important options that do not have direct analogs in the configuration file:

\begin{itemize}
 \item {\tt -{}-config <value>}: the main obligatory option to provide path to the configuration file;
 \item {\tt -{}-parallel}: set this option if multiple {\tt FIRE} instances are to be run at the same time, so that they do not conflict over temporary files;
 \item {\tt -{}-quiet}, {\tt -{}-QUIET}: suppress most of the output, even more so with the latter one;
 \item {\tt -{}-calc\_options <value>}: comma-separated options to be passed to calc library with {\tt fuel::setOption}; might be useful for testing things such as {\tt Symbolica} factorization of denominators;
 \item {\tt -{}-variables <value>}: set variable values, see the \ref{sec:mod_rec} section;
 \item {\tt -{}-large\_variables}: indicates that variables passed can be large and require an external library to be handled;
 \item {\tt -{}-folders <value>}: indicates how tables should be saved in sub-folders based on the quotient of the largest index shifts in passed variables and the value;
\end{itemize}

\subsection{Reconstruction tool and MPI parallelization}
The reconstruction binary has the following syntax: {\tt bin/reconstruct [options] filename range}. Here {\tt filename} should be the target reconstruction table which should follow the pattern where in the end of the filename the variable names or their values follow underscore-separated. The last number after the underscore stands for the prime number index or zero for the final reconstruction to rational numbers. Let us provide one example to demonstrate the syntax:

{\tt bin/reconstruct -{}-prime 1 -{}-method thiele \textbackslash \\-{}-reconstruction\_variable d\_100 \textbackslash \\tests/outputs/doublebox\_d\_7\_3\_1.tables 13}

Here a Thiele reconstruction by {\tt d} (for reconstruction methods, see section~\ref{sec:reconstruction}) is performed, taking $100$ as the initial value of {\tt d} and using up to $13$ tables. This is a double box example, where {\tt s} and {\tt t} have values $7$ and $3$, and the first large prime number is used as the modulus. The reconstruction tool will read the tables from disk with the same name as the target table with different values of {\tt substituted} from $100$ to $112$. If some of the tables are missing, it will try to reconstruct with existing ones (however, for the Zippel method, missing tables are not allowed due to the algorithm nature).

Calling {\tt bin/reconstruct -{}-help} prints help on all the options of the reconstruction tool. In most cases, the user is not supposed to call this tool directly, as it is called by the MPI wrapper (and the call syntax is seen in the log). We are not going into more details about this binary.

The MPI wrapper {\tt bin/FIRE7\_MPI} should be called to start the automatic reduction and reconstruction process. Let us demonstrate the options for the Zippel approach in a command provided in the {\tt examples} folder and comment on all the options:

{\tt mpirun -np 4 bin/FIRE7\_MPI \\-Z -r -E -S -D -B -P 10 -c examples/doubleboxN}

\begin{itemize}
 \item {\tt mpirun -np 4 bin/FIRE7\_MPI} --- launches 4 instances;
 \item {\tt -Z} or {\tt --Zippel} --- uses the Zippel approach;
 \item {\tt -r} or {\tt --reconstruct} --- requests automatic reconstruction;
 \item {\tt -E} or {\tt --early\_abortion} --- make {\tt FIRE} immediately abort in case it detects that with the current limits it cannot reconstruct some expression;
 \item {\tt -S} or {\tt --last\_separated} --- instructs that the last variable is separated (factorized) in the denominators;
 \item {\tt -D} or {\tt --delete\_tables} --- tells {\tt FIRE} to delete intermediate tables;
 \item {\tt -B} or {\tt --big\_primes} --- makes {\tt FIRE} use big primes as starting values;
 \item {\tt -P 10} or {\tt --rational\_reconstruction\_limit 10} --- limits to $10$ primes.
\end{itemize}

This is enough for a quick start. Other options can be seen in the
{\tt bin/FIRE7\_MPI --help} command and are mostly related to
fine-tuning performance on clusters. We will be happy to assist with
such tuning for large-scale problems.

\subsection{New tools for manipulating IBP tables}
\label{sec:tools}
We have also developed a number of useful utilities for working with {\tt FIRE} tables.
All of them produce the list of options if called with the {\tt --help} flag, so let us briefly outline their usage. Some practical examples will be given in Section \ref{sec:toolExamples}.

\begin{itemize}
\item {\tt combine} --- the {\tt combine} utility shipped with FIRE is used to apply IBP tables to linear combinations of integrals and takes three filenames as arguments. It takes a linear combination of integrals from the first file, uses IBP tables from the file in the second file to replace each integral by the reduction result, combines the reduction results with the respective prefactors, and saves the result to an output file. This can be used to e.g. reduce a complete amplitude which is a sum of integrals.
 \item {\tt diff} --- the {\tt diff} utility shipped with FIRE7 accepts two arguments and compares two files in FIRE's table format. The reason for using {\tt bin/diff} instead of a system {\tt diff} is that the tables can be equivalent in an algebraic sense but not identical bitwise. Note that it is important to specify the list of variables. The utility returns no output if the two tables agree. For non-coinciding tables, though, a non-zero error code is returned (useful for Unix pipes and scripts).
 \item {\tt add} --- accepts two input tables and one output file as arguments. The reduction results of each integral from the two tables are added to be the new reduction result in the output table. The subtraction mode can be turned on with the {\tt -s} command line flag to compute the difference between reduction results instead. This is more flexible than {\tt diff} for two reasons. First, an actual difference table is produced, though an additional zero/nonzero boolean conclusion can also be printed out if one passes the {\tt --verbose} or {\tt -V} command line option. Second, The input tables can be produced with two ``incompatible'' configuration files, which may cause {\tt diff} to fail.
 \item {\tt tables2rules} --- converts {\tt FIRE} tables format into {\tt Mathematica} set of rules;
 \item {\tt substitute} --- takes one table, substitutes a number of variables into it, and saves it as a new table file; The output filename is generated automatically according to the requested substitutions.

   Alternatively, one can specify a configuration file with {\tt -c} in the command line and omit the table filename. The filename will be deduced from the configuration file, and substitutions must be specified for all variables. The substitution values will be prepended before {\tt .tables} in output filenames. This utility is most useful for testing purposes;.
\end{itemize}

\subsection{Usage with LiteRed}

Let us start this small section with a quick note that although the package {\tt LiteRed}~\cite{Lee:2012cn, Lee:2013mka, LiteRed2} is shipped with {\tt FIRE}, this is a standalone package by another author and should be cited separately if used. We keep shipping {\tt Litered 1.8} with {\tt FIRE7}, but to work with {\tt LiteRed2}, the latter software should be downloaded and installed independently.

The way {\tt FIRE} can be used with {\tt LiteRed} requires {\tt Wolfram Mathematica}. A basic setup is the following:

\begin{itemize}
 \item Create a {\tt start} file, save it, unload the {\tt Mathematica} kernel.
 \item Run a generation of {\tt LiteRed} data related to the diagram in use. In general cases, we recommend using {\tt LiteRed} detections of zero sectors and symmetries, including external (between different sectors) and also internal (inside a sector) symmetries. Quit the {\tt Mathematica} kernel before the next step.
 \item Run the part of {\tt FIRE} that transforms the information in {\tt LiteRed} folders to an {\tt lbases} file that can be referred to by the {\tt config} setup for the {\tt c++} reduction. At the same time, do not forget to save a new {\tt start}/{\tt sbases} file, as it stores the information regarding which sectors are mapped to other sectors by the {\tt LiteRed} symmetries.
\end{itemize}

Now one can perform the {\tt c++} reduction with the use of {\tt LiteRed} symmetries. For the syntax details, please refer to the {\tt FIRE6} paper \cite{FIRE6}.

\section{Examples and benchmarks}
\label{sec:exampleAndBenchmark}
\subsection{Examples for IBP table tools}
\label{sec:toolExamples}
We present a mini-tutorial based on one-loop box integrals, to illustrate the use of the new tools introduced in Section \ref{sec:tools}, but we will start by reminding users about the basic usage of FIRE. The various files mentioned can be found inside the {\tt examples} folder.

\subsubsection{basic usage}

We define the integral family in Mathematica and export a {\tt start} file, as in {\tt examples/box.nb}:
\begin{lstlisting}
SetDirectory[NotebookDirectory[] <> "../"];
Get["FIRE7.m"];
Internal = {k};
External = {p1, p2, p4};
Propagators = {-k^2, -(k+p1)^2, -(k+p1+p2)^2,
               -(k+p1+p2+p4)^2};
Replacements = {p1^2->0, p2^2->0, p4^2->0,
                p1 p2->-S/2, p2 p4->-T/2,
                p1 p4->(S + T)/2, S->1, T->1};
PrepareIBP[];
Prepare[AutoDetectRestrictions->True, LI->True];
SaveStart["examples/box"];
\end{lstlisting}
This creates the file {\tt box.start}, which contains information about the integral family in a format that can be read by the C++ version of FIRE used in the terminal. We would like to reduce the integrals
\begin{equation}
  \label{eq:boxToReduce}
  G_1(2,2,1,1), \quad G_1(2,1,1,1)
\end{equation}
to master integrals. We have arbitrarily assigned the ``problem number'' to be 1, and therefore wrote $G_1$ above. The integrals are written to {\tt examples/box.m} with the content:
\begin{lstlisting}
{
  {1,{2,2,1,1}},
  {1,{2,1,1,1}}
}
\end{lstlisting}
The two integrals are shown as lists of indices, preceded by the problem number 1. The configuration file is {\tt examples/box.config}:
\begin{lstlisting}
#compressor        none
#threads           1
#fthreads          4
#variables         d
#start
#folder            examples/
#problem           1 box.start
#integrals         box.m
#output            ../tests/outputs/box.tables
\end{lstlisting}
We run the shell command (from the directory {\tt fire/FIRE7}, as usual)
\begin{lstlisting}
bin/FIRE7 -c examples/box
\end{lstlisting}
which produces the IBP reduction table {\tt tests/outputs/box.tables}.

\subsubsection{converting tables to Mathematica format}
In previous versions of FIRE, the only way to load an IBP table into Mathematica is through the {\tt LoadTables} function provided by FIRE's Mathematica package. FIRE 7 introduces a new tool, {\tt tables2rules}, which directly works on the command line without running Mathematica. Running
\begin{lstlisting}
bin/tables2rules tests/outputs/box.tables\
    tests/outputs/box.tables.m
\end{lstlisting}
converts the table to Mathematica format and writes to the latter file specified in the command line.
\subsubsection{divide and conquer: split master mode}
Complicated IBP reduction tasks can be divided into smaller jobs each
keeping only a subset of master integrals. We illustrate the usage
with the box integrals. First, we save the list of master integrals,
\begin{equation}
  \label{eq:boxMasters}
  G_1(1,1,1,1), \quad G_1(1,0,1,0), \quad G_1(0,1,0,1)
\end{equation}
to the file {\tt examples/box.masters}:
\begin{lstlisting}
{
  {1,{1,1,1,1}},
  {1,{1,0,1,0}},
  {1,{0,1,0,1}}
}
\end{lstlisting}
Then we create a configuration file {\tt box\_split.config} for IBP reduction targeting only
the first master integral, using the {\tt \#masters} syntax explained
in Section \ref{sec:split}:
\begin{lstlisting}
#compressor        none
#threads           1
#fthreads          4
#variables         d
#start
#folder            examples/
#problem           1 box.start
#masters           |1-1|box.masters
#integrals         box.m
#output            ../tests/outputs/box.tables
\end{lstlisting}
Similarly, we create another configuration file {\tt box\_split\_part2.config} targeting the remaining master integrals,
i.e. the 2nd to 3rd ones:
\begin{lstlisting}
#compressor        none
#threads           1
#fthreads          4
#variables         d
#start
#folder            examples/
#problem           1 box.start
#masters           |2-3|box.masters
#integrals         box.m
#output            ../tests/outputs/box.tables
\end{lstlisting}

Running the shell commands
\begin{lstlisting}
bin/FIRE7 -c examples/box_split
bin/FIRE7 -c examples/box_split_part2
\end{lstlisting}
produces the two tables:\\
{\tt tests/outputs/box.1-1.tables} and {\tt
  tests/outputs/box.2-3.tables},\\
which only include the 1st master integral and the 2nd-3rd master
integrals, respectively. Note that the output file names specified by
the configuration files have been automatically modified to include
the information about the range of the master integrals
included.
\subsubsection{{\tt add} tables from split master mode}
We combine the above two tables with {\tt bin/add}, specifying ``d'' as the only variable involved in the tables.
\begin{lstlisting}
bin/add -v d tests/outputs/box.1-1.tables \
    tests/outputs/box.2-3.tables \
    tests/outputs/box.added.tables
\end{lstlisting}
The last argument above is the output file. To check agreement with the table obtained previously without the split-master mode, we compute their difference with the ``subtraction'' mode of {\tt bin/add}:
\begin{lstlisting}
bin/add -v d -sV tests/outputs/box.added.tables \
    tests/outputs/box.tables /dev/null
\end{lstlisting}
We have discarded the output by directing it to {\tt /dev/null}, while using the ``verbose'' {\tt -V} flag to print out a conclusion,\\
\begin{lstlisting}
Zero table written
\end{lstlisting}
which confirms that the two tables are consistent, as desired.

\subsubsection{{\tt combine} to reduce linear combinations of integrals}
Suppose we would like to reduce a linear combination of the integrals in Eq.~\eqref{eq:boxToReduce},
\begin{equation}
  \label{eq:boxCombination}
  (d+2) G_1(2,2,1,1) + d^2 G_1(2,1,1,1) \, .
\end{equation}
We create a file, {\tt examples/box.combinations}:
\begin{lstlisting}
{
  {G[1,{2,2,1,1}],d+2},
  {G[1,{2,1,1,1}], d^2}
}
\end{lstlisting}
Then we run {\tt bin/combine} to reduce the linear combination against the existing IBP table, {\tt tests/outputs/box.tables},
\begin{lstlisting}
bin/combine -v d examples/box.combinations \
    tests/outputs/box.tables \
    tests/outputs/box.reduced
\end{lstlisting}
We have again specified $d$ as the only variable, and the output has been written to {\tt tests/outputs/box.reduced}. The output file contains, in Mathematica notation, the coefficients for each of the three master integrals, and the content should be self-explanatory.
\subsubsection{Streamlined direct reduction of linear combinations}
If one wants to reduce a linear combination of a large number of integrals, reducing all these integrals in the above workflow will result in a large table, even though the final result may be rather short. To bypass the need for large intermediate files, we can directly reduce linear combination of integrals. To do this for the simple example Eq.~\eqref{eq:boxCombination}, we create an integral combination file {\tt examples/boxc.m},
\begin{lstlisting}
{
  {
    {d+2, {1,{2,2,1,1}}},
    {d^2, {1,{2,1,1,1}}}
  }
}
\end{lstlisting}
More generally, the file content can be a list of linear combinations and also ordinary single integrals---the file above is a list of only one combination, which explains why the file has one more nested level of brackets than {\tt box.combinations}.

Then we create a new configuration file {\tt examples/boxc.config} that differs from {\tt examples/box.config} only in specifying {\tt boxc.m} for the {\tt \#integrals} option and {\tt boxc.tables} as the output. Running
\begin{lstlisting}
bin/FIRE7 -c examples/boxc
\end{lstlisting}
produces the output file. Inspecting the file, e.g.\ by converting it to a more readable Mathematica format with
\begin{lstlisting}
bin/tables2rules tests/outputs/boxc.tables \
    tests/outputs/boxc.tables.m
\end{lstlisting}
shows that the linear combination of integrals is reduced to master integrals.
\subsubsection{{\tt substitute} numerical values for variables in tables}
Sometimes, it is necessary to turn an analytic IBP table into a numerical one before further use. For example, the following command substitutes an arbitrary value of 53 for the $d$ variable, modulo the 1st large prime tabulated for use with FIRE (see Section \ref{sec:mod_rec}):
\begin{lstlisting}
bin/substitute -c examples/box -v 53_1
\end{lstlisting}
With the above command, the {\tt substitute} tool reads the configuration file {\tt examples/box.config} and deduces that the analytic IBP table is located at {\tt tests/outputs/box.tables}, then performs the requested substitutions and saves the result to {\tt tests/outputs/box\_53\_1.tables}. Generally, when multiple variables are involved, their numerical values need to be all given in the command line, separated by underscore characters, with the last number $n$ specifying the modulus as the $n$-th FIRE prime.
\subsubsection{Verifying agreement with direct numerical run}
Let us check that the above numerical substitution result agrees with a table produced by a direct numerical IBP run. First, let us back up the substitution result,
\begin{lstlisting}
cp tests/outputs/box_53_1.tables \
    tests/outputs/box_53_1.tables.bak
\end{lstlisting}
Then we use {\tt FIRE7p} to perform the numerical IBP run at $d=53$ modulo the 1st FIRE prime,
\begin{lstlisting}
bin/FIRE7p -v 53_1 -c examples/box
\end{lstlisting}
Let us verify that the newly created (overwritten) {\tt box\_53\_1.tables} agrees with the backup of the previous file:
\begin{lstlisting}
bin/diff -v d tests/outputs/box_53_1.tables \
    tests/outputs/box_53_1.tables.bak
\end{lstlisting}
The two files agree, as {\tt bin/diff} exited without output.\footnote{As covered previously, in more general situations where two tables are produced from incompatible configuration file settings, e.g. different split-master ranges, a more flexible (but more resource-intensive) alternative to {\tt bin/diff} is {\tt bin/add -sV}.}
\subsection{Benchmarks for modular IBP reduction}
\label{sec:benchmarks}

\begin{figure}
  \centering
  \includegraphics[width=0.44\textwidth]{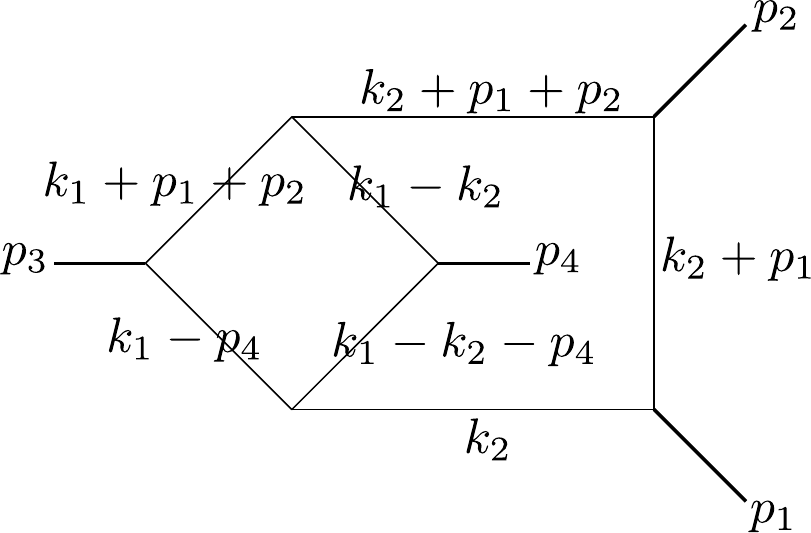}
  \caption{The nonplanar double box diagram with massive legs. The legs $p_1$, $p_2$ and $p_3$ share the same mass, while $p_4$ has a different mass. The list of seven propagator denominators and two ISPs are $(k_1-k_2-p_4)^2$, $(k_1+p_1+p_2)^2$, $(k_1-p_4)^2$, $(k_1-k_2)^2$, $k_2^2$, $(k_2+p_1)^2$, $(k_2+p_1+p_2)^2$, $(k_1+p_1)^2$, and $(k_2-p_4)^2$.}
  \label{fig:offshell_nonplanar_dbox}
\end{figure}

\begin{figure}
  \centering
  \includegraphics[width=0.4\textwidth]{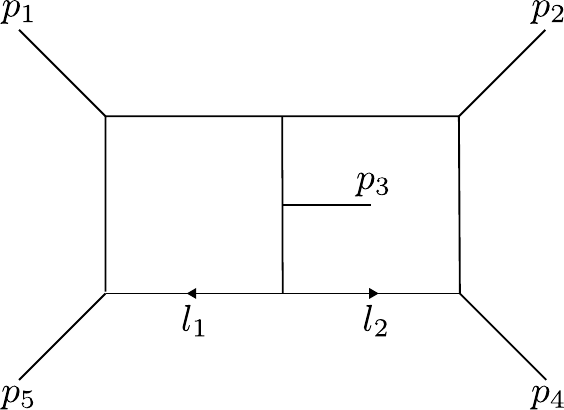}
  \caption{The nonplanar double pentagon diagram. The list of eight propagator denominators and three ISPs are $l_1^2$, $(l_1-p_5)^2$, $(l_1-p_5-p_1)^2$, $(l_2-p_4-p_2)^2$, $(l_2-p_4)^2$, $(l_2)^2$, $(l_1+l_2)^2$, $(l_1+l_2+p_3)^2$, $(l_2+p_5)^2$, $(l_1+p_4)^2$, and $(l_1+p_4+p_2)^2$.}
  \label{fig:nonplanarDoublePentagon}
\end{figure}

\begin{figure}
  \centering
  \includegraphics[width=0.6\textwidth]{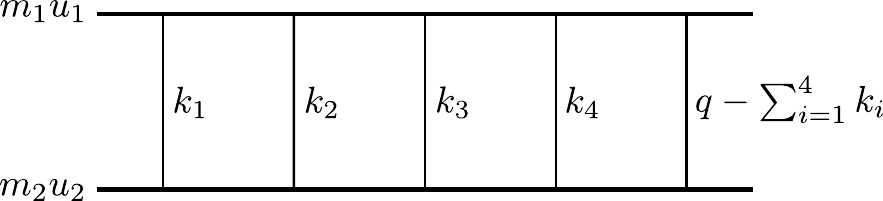}
  \caption{The soft-expanded 4-loop quadruple box diagram. The massless vertical lines in the middle are considered as soft. The expansion linearizes the massive propagators shown as horizontal lines. The massive momenta running through the top horizontal line and bottom horizontal line are approximately $m_1 u_1$ and $m_2 u_2$, respectively, where $u_1$ and $u_2$ are normalized velocities. The thirteen propagators are nine ISPs are $-2 k_1 \cdot u_2$, $-2(k_1+k_2) \cdot u_2$, $-2(k_1+k_2+k_3) \cdot u_2$, $-2(k_1+k_2+k_3+k_4) \cdot u_2$, $2 k_1 \cdot u_1$, $2(k_1+k_2) \cdot u_1$, $2(k_1+k_2+k_3) \cdot u_1$, $2(k_1+k_2+k_3+k_4) \cdot u_1$, $k_1^2$, $k_2^2$, $k_3^2$, $k_4^2$, $(q-k_1-k_2-k_3-k_4)^2$, $k_1 \cdot q$, $k_2 \cdot q$, $k_3 \cdot q$, $k_1 \cdot k_2$, $k_1 \cdot k_3$, $k_1 \cdot k_4$, $k_2 \cdot k_3$, $k_2 \cdot k_4$, and $k_3 \cdot k_4$.}
  \label{fig:soft_quad_box}
\end{figure}

We provide benchmarks comparing the modular arithmetic performance of
FIRE 6.5 and FIRE 7 with different orderings (default, Aia, Apn), as
explained in Section \ref{sec:orderings}, for the IBP reduction of the
following integrals, setting all kinematic variables to random
numerical values and performing the IBP reduction modulo a 64-bit
prime number. The multi-threading settings {\tt \#threads} and {\tt
  \#fthreads} (as explained in Section \ref{sec:config} are both 4 in
the runs.
\begin{itemize}
\item The nonplanar double box diagram with four massive legs, three
  of which share the same mass, and massless internal lines, with
  $p_1^2=p_2^2=p_3^2 \neq p_4^2$, in Fig.\
  \ref{fig:offshell_nonplanar_dbox}, with two tensor integrals each
  with 8 ISP powers, $[(k_1+p_1)^2]^4 [(k_2-p_4)^2]^4$ and
  $[(k_1+p_1)^2]^8$.
\item The nonplanar double pentagon diagram Fig.\
  \ref{fig:nonplanarDoublePentagon}, with the tensor integral with 4
  ISP powers, $(l_2+p_5)^2 (l_1+p_4)^2 [(l_1+p_2+p_4)^2]^2$.
\item The soft-expanded 4-loop quadruple box diagram Fig.\
  \ref{fig:soft_quad_box} for the scattering of two massive particles
  with a soft momentum transfer in the t-channel, with the tensor
  integral with 3 ISP powers, $(k_2\cdot k_4)(k_3\cdot k_4)^2$. This
  diagram is taken from a 4-loop calculation of the classical limit of
  charged particle scattering in electrodynamics \cite{Bern:2023ccb}.
\end{itemize}
The benchmark results, including run times and estimated memory usage,
are in Table \ref{tab:benchmarks}. The machine used has the Intel Xeon
Platinum 8280 CPU and 192 GB of DDR4 RAM. We can see that FIRE 7,
under any of the orderings chosen, outperforms FIRE 6.5 under the
default ordering. The results for the fastest ordering for FIRE 7 are
highlighted in red, while FIRE 6.5 results are highlighted in blue, to
clearly highlight the ordering choice with the largest time
savings. Furthermore, we test the multiprime mode, explained in
Section \ref{sec:multiprime}, to perform IBP reduction for 16
numerical probes simultaneously while reusing the best ordering choice
found for each integral family.\footnote{Note that the ordering is
  also affected by how the propagators and ISPs are ordered by the
  user, and this information is given in the figure captions of the
  diagrams.} The \emph{average} run time per numerical probe and the
memory usage is shown in the last column of Table
\ref{tab:benchmarks}. Compared with FIRE 6.5 with default orderings,
the multiprime mode saves the run time by a factor of about 5 to 50
depending on the integral family, though sometimes with higher memory
usage.
\begin{table}[ht]
  \centering
  {
    \footnotesize
    \begin{tabular}{|c|c|c|c|c|c|}
      \hline
      Integral family & FIRE 6.5 & default & Aia & Apn & Multiprime \\ \hline
      \makecell{Double box\\ with massive legs} & \textcolor{blue}{52s, 0.45G} & 43s, 0.46G & 45s, 0.43G & \textcolor{red}{39s, 0.39G} & \textcolor{OliveGreen}{11s, 1.4G} \\ \hline
      \makecell{nonplanar\\ double pentagon} & \textcolor{blue}{100s, 0.38G} & 92s, 0.41G & \textcolor{red}{60s, 0.38G}  & 78s, 0.40G & \textcolor{OliveGreen}{20s, 1.6G} \\ \hline
      \makecell{soft-expanded\\ 4-loop box} & \textcolor{blue}{237s, 6.4G} & 151s, 6.9G & \textcolor{red}{30s, 3.7G} & 152s, 6.4G & \textcolor{OliveGreen}{4.7s, 4.1G} \\ \hline
    \end{tabular}
  }
  \caption{Benchmark results for three integral families, as run times
    in seconds and estimated memory usage in Gigabytes, with the
    modular arithmetic mode of FIRE. Each benchmark is run three
    times, with the shortest time taken to reduce noise from
    background processes.}
  \label{tab:benchmarks}
\end{table}

\subsection{Benchmark for analytic IBP reduction}
\label{sec:benchmarksAnalytic}
\begin{figure}[h]
  \centering
  \includegraphics[width=0.44\textwidth]{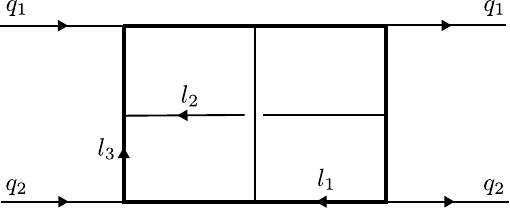}
  \caption{The three-loop forward-scattering diagram with massive internal lines, with $q_1^2=q_2^2=0$, $(q_1+q_2)^2 = s$. The list of 10 propagators are $l_3^2 - m^2$, $(l_2+l_3)^2 - m^2$, $l_1^2 - m^2$, $(l_3-q_2)^2 - m^2$, $(l_2+l_3+q_1)^2 - m^2$, $(l_1+q_2)^2-m^2$, $(l_1+l_2+q_1+q_2)^2-m^2$, $(l_1+l_2+q_2)^2 - m^2$, $l_2^2$, and $(l_1-l_3+q_2)^2$. The two ISPs are $(l_1+l_2+l_3)^2$ and $(l_1-q_1)^2$.}
  \label{fig:threeLoopWindow}
\end{figure}

We also present a benchmark for analytic IBP reduction for a 3-loop integral family with massive internal lines. The kinematics is exact forward scattering with massless incoming momenta, $q_1 + q_2 \rightarrow q_1 + q_2$. The diagram is shown in Fig~\ref{fig:threeLoopWindow}. We reduce only one sample integral in this benchmark, where the 4th propagator $(l_3-q_2)^2 - m^2$ is raised to a 2nd power, the 8th propagator is absent (canceled), and the numerator is the product of the two ISPs, $(l_1+l_2+l_3)(l_1-q_1)^2$. FLINT (default for FIRE 7) is used as the simplifier for all the runs.
The performance of FIRE 6.5 and FIRE 7 with default orderings, as well as FIRE 7 with Aia and Apn orderings, are shown in Table \ref{tab:benchmarksAnalytic}. The various settings for running the benchmarks are identical to Section \ref{sec:benchmarks}, except that {\tt \#threads} and {\tt \#fthreads} are now set to 24.
\begin{table}[ht]
  \centering
  {
    \footnotesize
    \begin{tabular}{|c|c|c|c|c|c|}
      \hline
      FIRE 6.5 & default & Aia & Apn \\ \hline
      \textcolor{blue}{$11.6 \times 10^3$ s, 142G} & \textcolor{red}{$4.7 \times 10^3$ s, 67 G} & $5.4 \times 10^3$s, 78G & $12.1 \times 10^3$s, 211G \\ \hline
    \end{tabular}
  }
  \caption{Benchmark results for the three-loop forward scattering
    diagram with massive internal lines, as run times in seconds and
    estimated memory usage in Gigabytes, with the analytic mode of
    FIRE.}
  \label{tab:benchmarksAnalytic}
\end{table}

This time, FIRE 7 with the default ordering (also dependent on the order in which we supplied the list of propagators and ISPs) achieved the best performance. The results here are complementary to the ones for modular reduction in Section \ref{sec:benchmarks}, and demonstrate the advantage of FIRE 7 for larger-scale analytic reduction runs.

\section{Conclusion}

We have presented the new version 7 of {\tt FIRE}, a program for
integration-by-parts reduction of Feynman integrals, aimed at
cutting-edge calculations in perturbative quantum field theory. A main
change is enhanced performance for IBP reduction with modular
arithmetic, achieved by an improved presolve of IBP identities before
replacing symbolic indices (propagator powers) by integer numerical
indices. The new presolve involves Gaussian elimination with both
forward and backward stages, and reduces the complexity of the IBP
identities before proceeding with a conventional Laporta-style seeding
of IBP equations. The user can also adjust the ordering of integrals
in the implementation of the Laporta algorithm, which can have
significant impacts on run times and memory usage, as shown in the
benchmarks. Note that experimenting with ordering settings is
particularly suited to the modular approach, since one can find the
best ordering setting using IBP runs at a small number of numerical
probe points, and the performance characteristics will remain
unchanged at a large number of new probe points needed to complete the
reconstruction of analytic results.

The multiprime mode offers another significant performance
enhancement, by performing IBP reduction at multiple probe points
simultaneously to reduce the overhead incurred per point. In certain
cases, combining a suitable ordering choice and the multiprime mode
has resulted in a 50-fold speedup over the previous version of FIRE.

The usability of modular IBP reduction is greatly improved by the
complete automation of MPI-parallelized reduction and subsequent
reconstruction of coefficients. The performance of analytic IBP
reduction is also enhanced, benefiting from improvements such as the
updated presolve. Several new command-line tools are introduced for
efficiently manipulating IBP tables, improving the user experience
beyond the core IBP reduction tasks.

This release brings most of the current code of {\tt FIRE} for public
usage.  An experimental implementation of improved seeding
\cite{Bern:2024adl, Driesse:2024xad, Guan:2024byi,
  lange2025kira3integralreduction,
  vonHippel:2025okr,Song:2025pwy,Zeng:2025xbh} remains in development
and has been applied in Ref.~\cite{Bern:2025zno}, to be released in
the future.

\section{Acknowledgments}
The work of AS was supported by the Ministry of Education and Science of the Russian Federation as part of the program of the Moscow Center for Fundamental and Applied
Mathematics under Agreement No.\ 075-15-2025-345 (in part of developing reconstruction algorithms).
The study of AS was conducted under the state assignment of Lomonosov Moscow State University (in part of optimizing reduction within the polynomial approach).
M.Z.’s work is supported in part by the U.K.\ Royal Society through Grant URF\textbackslash R1\textbackslash 20109. For the purpose of open access, the authors have applied a Creative Commons Attribution (CC BY) license to any Author Accepted Manuscript version arising from this submission.
The authors acknowledge the Texas Advanced Computing Center (TACC) at the University of Texas at Austin for providing high-performance computing resources that have contributed to the research results reported within this paper. The research was carried out using the equipment of the shared research facilities of HPC computing resources at Lomonosov Moscow State University \cite{voevodin2019lomonosov2}.

\bibliographystyle{elsarticle-num-names}
\bibliography{fire7}
\end{document}